\documentclass[doublecol,figures]{epl2}
\usepackage{amsmath,amssymb}

\title{Empirical analysis of web-based user-object bipartite networks}
\author{Ming-Sheng Shang$^1$ \and Linyuan L\"u$^2$ \and Yi-Cheng Zhang$^{1,2}$ \and Tao
Zhou$^{2,3}$\footnote{Corresponding author: zhutou@ustc.edu}}
\shortauthor{Ming-Sheng Shang, Linyuan L\"u, Yi-Cheng Zhang, Tao
Zhou} \institute{$^1$Web Science Center, School of Computer Science
and Engineering, University of Electronic Science and Technology,
610054 Chengdu, P. R. China\\ $^2$Department of Physics, University
of Fribourg, CH-1700~Fribourg, Switzerland
\\ $^3$Department of Modern Physics, University of Science and Technology of China, Hefei
230026, P. R. China}


\abstract{Understanding the structure and evolution of web-based
user-object networks is a significant task since they play a crucial
role in e-commerce nowadays. This Letter reports the empirical
analysis on two large-scale web sites, \emph{audioscrobbler.com} and
\emph{del.icio.us}, where users are connected with music groups and
bookmarks, respectively. The degree distributions and degree-degree
correlations for both users and objects are reported. We propose a
new index, named \emph{collaborative clustering coefficient}, to
quantify the clustering behavior based on the collaborative
selection. Accordingly, the clustering properties and
clustering-degree correlations are investigated. We report some
novel phenomena well characterizing the selection mechanism of web
users and outline the relevance of these phenomena to the
information recommendation problem.}
\begin{document}
\maketitle

\section{Introduction}
The last decade has witnessed tremendous activities devoted to the
understanding of complex networks
\cite{Albert2002,Dorogovtsev2002,Newman2003,Boccaletti2006,Costa2007}.
A particular class of networks is the \emph{bipartite networks},
whose nodes are divided into two sets $X$ and $Y$, and only the
connection between two nodes in different sets is allowed. Many
systems are naturally modeled as bipartite networks
\cite{Holme2003}: the human sexual network \cite{Liljeros2001}
consists of men and women, the metabolic network \cite{Jeong2000}
consists of chemical substances and chemical reactions, the
collaboration network \cite{Zhang2006} consists of acts and actors,
the Internet telephone network consists of personal computers and
phone numbers \cite{Xuan2009}, etc. In addition to the empirical
analysis on the above-mentioned bipartite networks, great effort has
been made in how to characterize bipartite networks
\cite{Lind2005,Estrada2005,Peltomaki2006}, how to project bipartite
networks into monopartite networks
\cite{Lambiotte2005,Zhou2007,Wang2009} and how to model bipartite
networks \cite{Ramasco2004,Ohkubo2005,Goldstein2005,Guillaume2006}.

\begin{figure}
\begin{center}
\includegraphics[width=6cm]{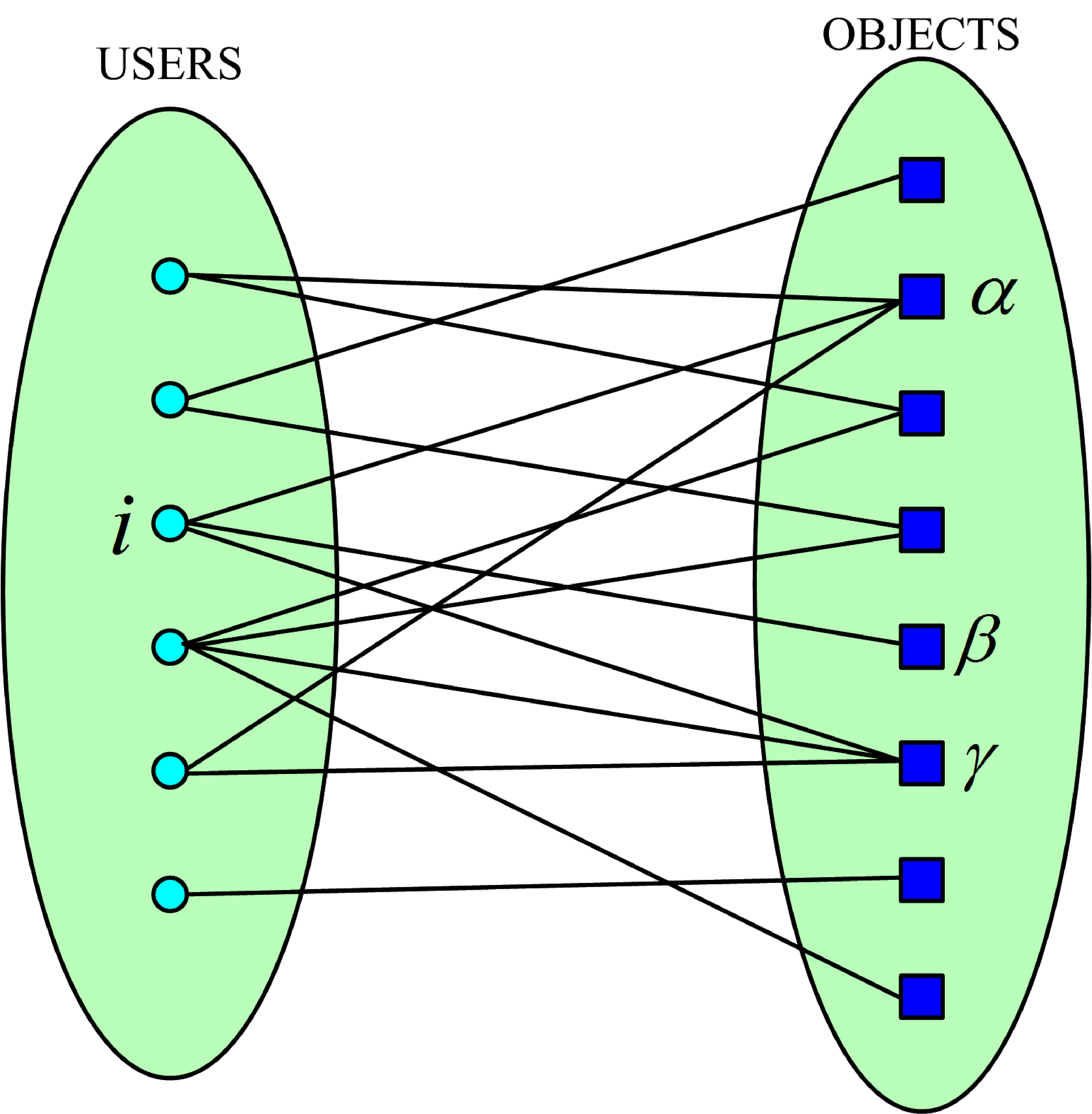}
\caption{(Color online) Illustration of a small user-object
bipartite network.}
\end{center}
\end{figure}

\begin{table*}
\caption{The basic properties of the two data sets. $N$, $M$ and $E$
denote the numbers of users, objects and edges, respectively.
$\langle k\rangle$ and $\langle d\rangle$ are the average user
degree and average object degree. $C_u$ and $C_o$ are the
collaborative clustering coefficients for users and objects, and for
comparison, $\bar{s_o}$ and $\bar{s_u}$ are the average similarities
over all object pairs and over all user pairs, respectively. The
user selection is considered to be highly clustered since $C_u\gg
\bar{s_o}$.}
\begin{center}
\begin{tabular} {cccccccccc}
\hline \hline
Data     & $N$  &  $M$  & $E$ & $\langle k\rangle$ & $\langle d\rangle$ & $C_u$ & $\bar{s_o}$ & $C_o$ & $\bar{s_u}$ \\
\hline
Audioscrobbler.com     & 35916 &  617900  & 5028580 & 140.01 & 8.14 & 0.0267 & $9.96 \times 10^{-5}$ & 0.0198 & $4.82\times 10^{-3}$ \\
Del.icio.us     & 10000  &  232658  & 1233995 & 123.40 & 5.30 & 0.0338 & $4.64\times 10^{-4}$ & 0.0055 & $8.10 \times 10^{-4}$ \\
\hline \hline
\end{tabular}
\end{center}
\end{table*}

An important class of bipartite networks is the \emph{web-based
user-object networks}, which play the central role in e-commerce for
many online selling sites and online services sites
\cite{Schafer2001}. This class of networks has two specific evolving
mechanisms different from the well-understood act-actor bipartite
networks and human sexual networks. Firstly, connections between
existent users and objects are generated moment by moment while this
does not happen in act-actor networks (e.g., one can not add authors
to a scientific paper after its publication). Secondly, users are
active (to select) while objects are passive (to be selected). This
is different from the human sexual networks where in principle both
men and women are active. In a word, the user-object networks are
driven by selection of users while the human sexual networks are
driven by matches. Bianconi \emph{et al.} \cite{Bianconi2004}
investigated the effects of the selection mechanisms of users on the
network evolution. Lambiotte and Ausloos
\cite{Lambiotte2005b,Lambiotte2006} analyzed the web-based bipartite
network consisted of listeners and music groups, especially, they
developed a percolation-based method to uncover the social
communities and music genres. Zhou \emph{et al.} \cite{Zhou2007}
proposed a method to better measure the user similarity in general
user-object bipartite networks, which has found its applications in
personalized recommendations. Huang \emph{et al.} \cite{Huang2007}
analyzed the user-object networks (called \emph{consumer-product
networks} in Ref. \cite{Huang2007}) to better understand the
purchase behavior in e-commerce settings\footnote{Instead of the
direct analysis on bipartite networks, Huang \emph{et al.}
\cite{Huang2007} concentrated on the monopartite networks obtained
from the bipartite networks.}. Gruji\'c \emph{et al.}
\cite{Grujic2008,Grujic2009} studied the clustering patterns and
degree correlations of user-movie bipartite networks according to
the large-scale Internet Movie Database (IMDb), and applied a
spectral analysis method to detect communities in the projected
weighted networks. They found the monopartite networks for both
users and movies exhibit an assortative behavior while the bipartite
network shows a disassortative mixing pattern.

This Letter reports the empirical analysis on two well-known web
sites, \emph{audioscrobbler.com} and \emph{del.icio.us}, where users
are connected with music groups and bookmarks, respectively. Our
main findings are threefold: (i) All the object-degree distributions
are power-law, while the user-degree distributions obey stretched
exponential functions. (ii) The networks exhibit disassortative
mixing patterns, indicating that the fresh users tend to view
popular objects and the unpopular objects are usually collected by
very active users. (iii) We propose a new index, named
\emph{collaborative clustering coefficient}, to quantify the
clustering behavior based on the collaborative connections. The two
networks are of high collaborative clustering coefficients for both
users and objects. For the lower-degree objects, a negative
correlation between the object collaborative clustering coefficient
and the object degree is observed, which disappears when the degree
exceeds the average object degree. For audioscrobbler.com, the user
collaborative clustering coefficient is strongly negatively
correlated with the user degree, decaying in an exponential form for
low degrees.

\begin{figure}
\begin{center}
\includegraphics[width=7cm]{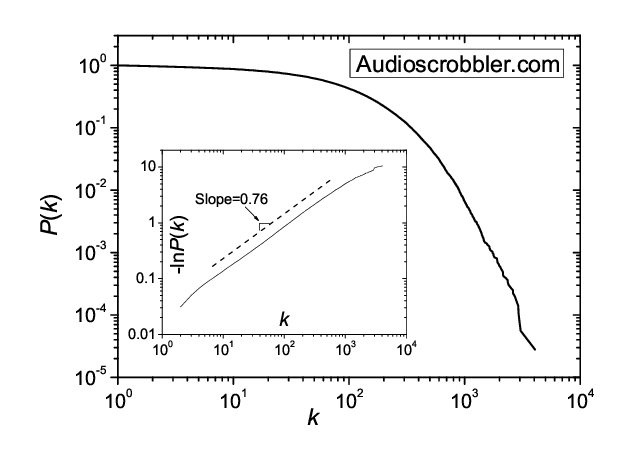}
\includegraphics[width=7cm]{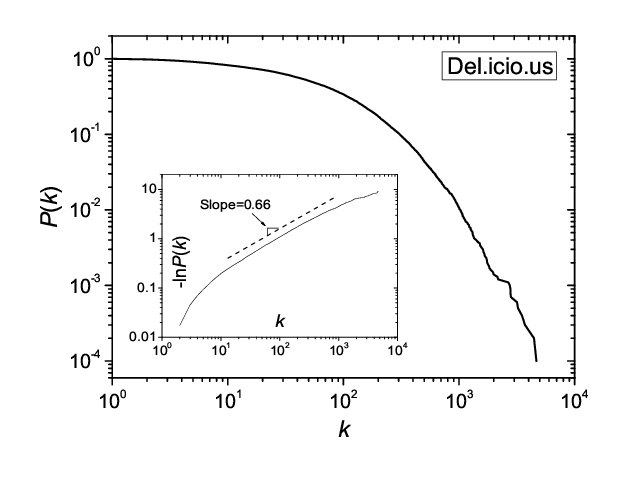}
\caption{Distributions of user degrees, which obey the stretched
exponential form \cite{Laherrere1998,Zhou2007b}. We therefore plot
the cumulative distribution $P(k)$ instead of $p(k)$ and show the
linear fittings of $\texttt{log}(-\texttt{log}P(k))$ vs.
$\texttt{log}k$ in the insets. }
\end{center}
\end{figure}

\begin{figure}
\begin{center}
\includegraphics[width=6cm]{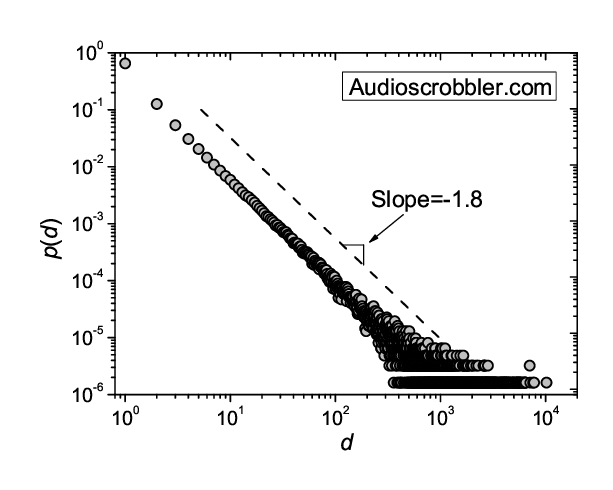}
\includegraphics[width=6.4cm]{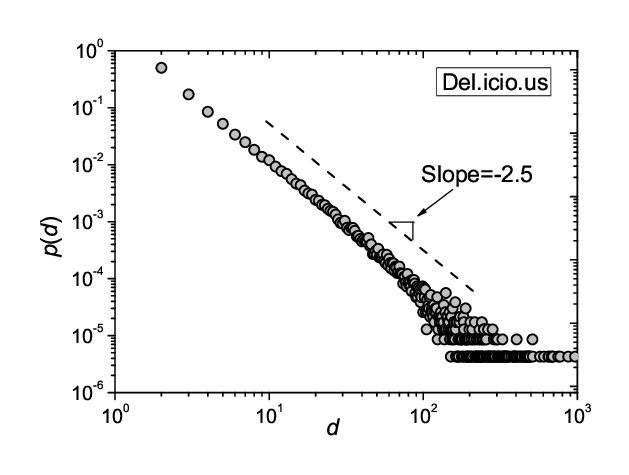}
\caption{Distributions of object degrees, which are power-law (they
can pass the Kolmogorov-Smirnov test with threshold quantile 0.9)
with exponents obtained by using the maximum likelihood estimation
\cite{Goldstein2004}. }
\end{center}
\end{figure}

\section{Basic Concepts}
Figure 1 illustrates a small bipartite network that consists of six
users and eight objects. The degree of user $i$, denoted by $k_i$,
is defined as the number of objects connected to $i$. Analogously,
the degree of object $\alpha$, denoted by $d_\alpha$, is the number
of users connected to $\alpha$. For example, as shown in Fig. 1,
$k_i=d_\alpha=3$. The density function, $p(k)$, is the probability
that a randomly selected user is of degree $k$, while the cumulative
function, $P(k)$, denotes the probability that a randomly selected
user is of degree no less than $k$. The \emph{nearest neighbors'
degree} for user $i$, denoted by $d_{\texttt{nn}}(i)$, is defined as
the average degree over all the objects connected to $i$. For
example, as shown in Fig. 1,
$d_\texttt{nn}(i)=\frac{d_\alpha+d_\beta+d_\gamma}{3}=\frac{7}{3}$.
The degree-dependent nearest neighbors' degree, $d_{\texttt{nn}}(k)$
is the average nearest neighbors' degree over all the users of
degree $k$, that is, $d_\texttt{nn}(k)=\langle
d_\texttt{nn}(i)\rangle_{k_i=k}$. Corresponding definitions for
objects, say $p(d)$, $P(d)$, $k_{\texttt{nn}}(\alpha)$ and
$k_\texttt{nn}(d)$, are similar and thus omitted here.

The traditional clustering coefficient \cite{Watts1998} cannot be
used to quantify the clustering pattern of a bipartite network since
it always give a zero value. Lind \emph{et al.} \cite{Lind2005}
proposed a variant counting the rectangular relations instead of
triadic clustering, which can be applied to general bipartite
networks. However, this Letter aims at a special class of bipartite
networks, and thus we propose a new index to better characterize the
clustering patterns resulted from the collaborative interests of
users. A standard measure of object similarity according to the
collaborative selection is the \emph{Jaccard similarity}
\cite{Jaccard1901}, $s_{\alpha\beta}=\frac{|\Gamma_\alpha\bigcap
\Gamma_\beta|}{|\Gamma_\alpha\bigcup \Gamma_\beta|}$, where
$\Gamma_\alpha$ and $\Gamma_\beta$ are the sets of neighboring nodes
of $\alpha$ and $\beta$, respectively. Obviously,
$s_{\alpha\beta}=s_{\beta\alpha}$ and $0\leq s_{\alpha\beta} \leq 1$
for any $\alpha$ and $\beta$. For example, as shown in Fig. 1,
$s_{\alpha\beta}=s_{\beta\gamma}=\frac{1}{3}$ and
$s_{\alpha\gamma}=\frac{1}{2}$. The \emph{collaborative clustering
coefficient} of user $i$ is then defined as the average similarity
between $i$'s selected objects:
$C_u(i)=\frac{1}{k_i(k_i-1)}\sum_{\alpha\neq \beta}s_{\alpha\beta}$,
where $\alpha$ and $\beta$ run over all $i$'s neighboring objects.
For example, as shown in Fig. 1, the collaborative clustering
coefficient of user $i$ is $C_u(i)=\frac{7}{18}$. The user
collaborative clustering coefficient of the whole network is defined
as $C_u=\frac{1}{N'}\sum_iC_u(i)$, where $i$ runs over all users
with degrees larger than 1 and $N'$ denotes the number of these
users. The degree-dependent collaborative clustering coefficient,
$C_u(k)$, is defined as the average collaborative clustering
coefficient over all the $k$-degree users. Corresponding definitions
for objects are as following: (i)
$C_o(\alpha)=\frac{1}{d_\alpha(d_\alpha-1)}\sum_{i\neq j}s_{ij}$,
where $s_{ij}=\frac{|\Gamma_i\bigcap \Gamma_j|}{|\Gamma_i\bigcup
\Gamma_j|}$ is the Jaccard similarity between users $i$ and $j$;
(ii) $C_o=\frac{1}{M'}\sum_\alpha C_o(\alpha)$, where $M'$ denotes
the number of objects with degrees larger than 1; (iii) $C_o(d)$ is
the average collaborative clustering coefficient over all the
$d$-degree objects.

\section{Data}
This Letter analyzes two data sets. One is downloaded from
audioscrobbler.com\footnote{Audioscrobbler.com is a well-known
collaborative filtering web site that allows user to create the
personal web pages as their music libraries and to discover new
music groups form other users' libraries.} in January 2005 by
Lambiotte and Ausloos \cite{Lambiotte2005b,Lambiotte2006}, which
consists of a listing of users, together with the list of music
groups the users own in their libraries. Detailed information about
this data set can be found in Refs.
\cite{Lambiotte2005b,Lambiotte2006}. The other is a random sampling
of $10^4$ users together with their collected bookmarks (URLs) from
del.icio.us\footnote{Del.icio.us is one of the most popular social
bookmarking web sites, which allows users not only to store and
organize personal bookmarks, but also to look into other users'
collections and find what they might be interested in.} in May 2008
\cite{Zhang2009}. Table 1 summarizes the basic statistics of these
two data sets.

\begin{figure}
\begin{center}
\includegraphics[width=5.8cm]{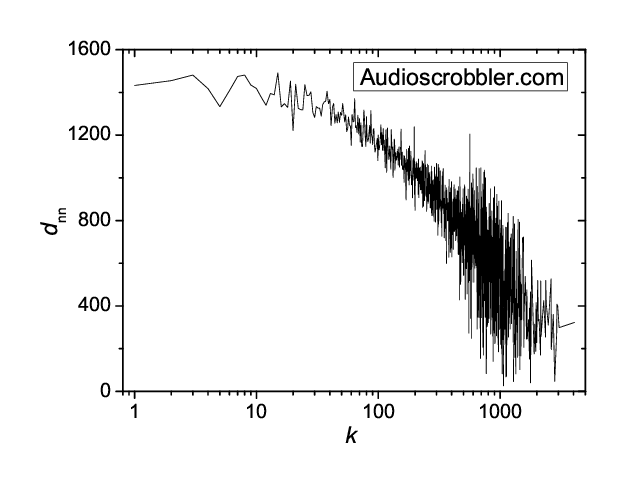}
\includegraphics[width=5.8cm]{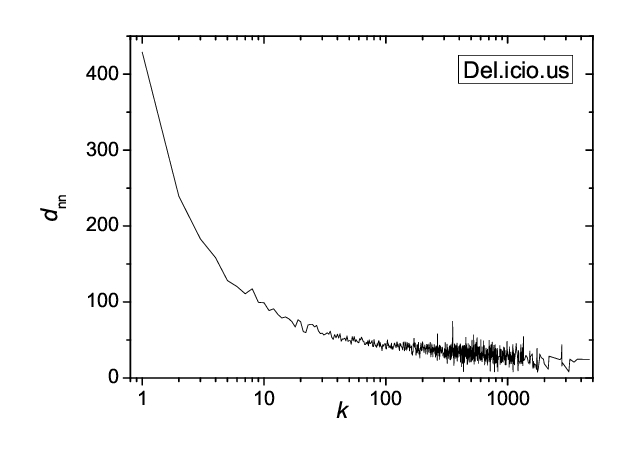}
\caption{The degree-dependent nearest neighbors' degree,
$d_\texttt{nn}(k)$, as a function of user-degree, $k$.}
\end{center}
\end{figure}

\begin{figure}
\begin{center}
\includegraphics[width=5.8cm]{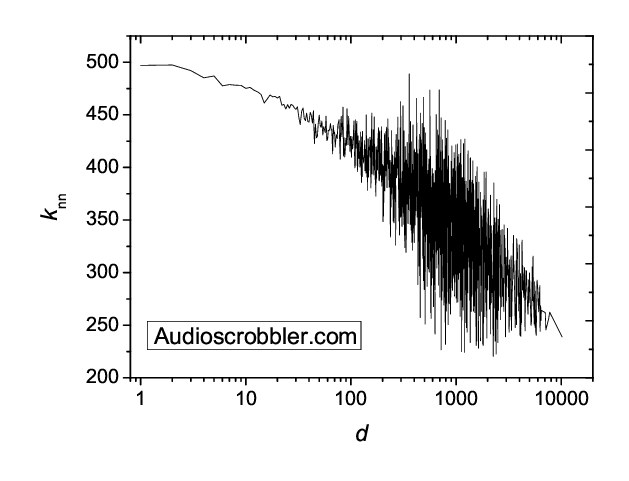}
\includegraphics[width=6.0cm]{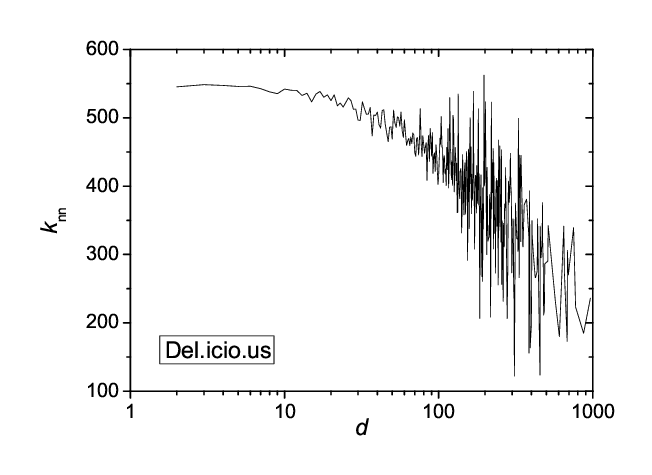}
\caption{The degree-dependent nearest neighbors' degree,
$k_\texttt{nn}(d)$, as a function of object-degree, $d$.}
\end{center}
\end{figure}

\section{Empirical Results}
Figure 2 reports the degree distributions for users, which do not
follow either the power-law form or the exponential form. In fact,
they lie in between exponential and power-law forms, and can be well
fitted by the so-called \emph{stretched exponential distributions}
\cite{Laherrere1998,Zhou2007b}, as $p(k)\sim
k^{\mu-1}\texttt{exp}\left[ -(\frac{k}{k_0})^\mu\right]$, where
$k_0$ is a constant and $0\leq\mu\leq1$ is the characteristic
exponent. The borderline $\mu=1$ corresponds to the usual
exponential distribution. For $\mu$ smaller than one, the
distribution presents a clear curvature in a log-log plot. The
exponent $\mu$ can be determined by considering the cumulative
distribution $P(k)\sim \texttt{exp}\left[
-(\frac{k}{k_0})^\mu\right]$, which can be rewritten as
$\texttt{log}(-\texttt{log}P(k))\sim \mu \texttt{log}k$. Therefore,
Using $\texttt{log}k$ as $x$-axis and
$\texttt{log}(-\texttt{log}P(k))$ as $y$-axis, if the corresponding
curve can be well fitted by a straight line, then the slope equals
$\mu$. Accordingly, as shown in Fig. 2, the exponents $\mu$ for
audioscrobbler.com and del.icio.us are 0.76 and 0.66 respectively.
These results have refined the previous statistics
\cite{Lambiotte2005b}, where the exponential function is directly
used to fit the user degree distribution of audioscrobbler.com. As
shown in Fig. 3, all the object-degree distributions are power laws,
as $p(d)\sim d^{-\phi}$. The exponents, $\phi$, obtained by the
maximum likelihood estimation \cite{Goldstein2004}, are shown in the
corresponding figures.

\begin{figure}
\begin{center}
\includegraphics[width=7cm]{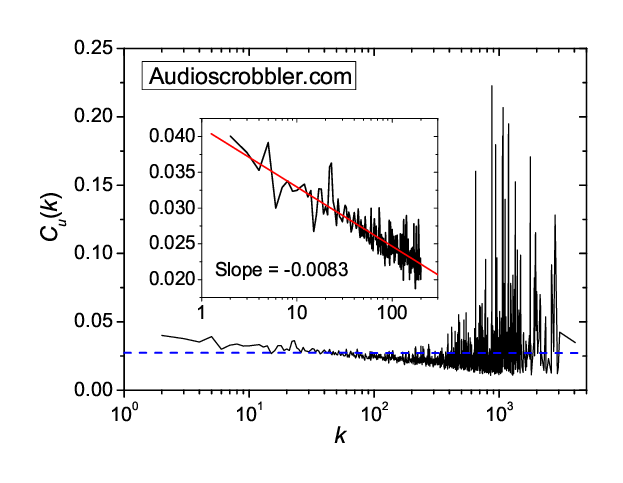}
\includegraphics[width=7cm]{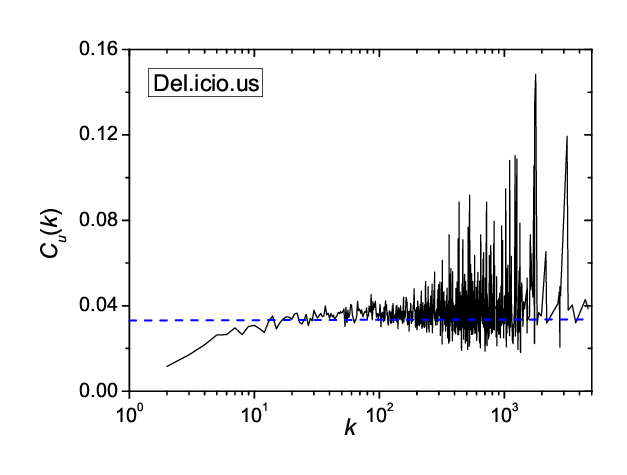}
\caption{(Color online) The clustering-degree correlations for
users. Blue dash lines denote the collaborative clustering
coefficients of the whole networks, $C_u$. The inset displays the
early decaying behavior of $C_u(k)$ for audioscrobbler.com, which
can be well fitted by an exponential form as $C_u(k)\sim
\texttt{e}^{-0.0083k}$.}
\end{center}
\end{figure}

As shown in Fig. 4 and Fig. 5, for both users and objects, the
degree is negatively correlated with the average nearest neighbors'
degree, exhibiting a disassortative mixing pattern. This result is
in accordance with the user-movie bipartite network
\cite{Grujic2008,Grujic2009}, indicating that the fresh users tend
to view popular objects and the unpopular objects are usually
collected by very active users. The correlation between
$d_\texttt{nn}$ and $k$ is stronger than this between
$k_\texttt{nn}$ and $d$, which may be caused by the fact that the
users are active while the objects are passive.

\begin{figure}
\begin{center}
\includegraphics[width=7.5cm]{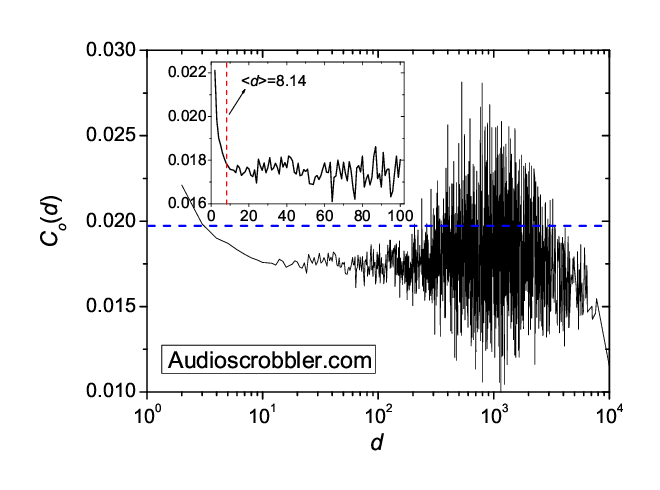}
\includegraphics[width=7.5cm]{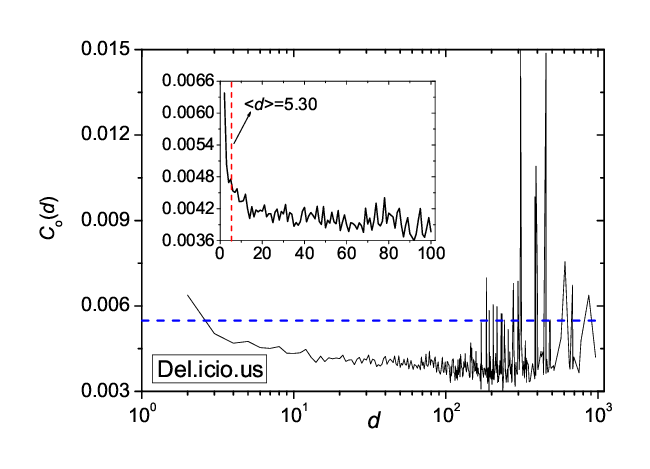}
\caption{(Color online) The clustering-degree correlations for
objects. Blue dash lines denote the collaborative clustering
coefficients of the whole networks, $C_o$. The insets display the
early decaying behavior of $C_o(d)$, with the read dash lines
denoting the average object degrees.}
\end{center}
\end{figure}

Table 1 reports the user collaborative clustering coefficients and
object collaborative clustering coefficients for the whole networks.
For comparison, we calculate the average user similarity over all
user pairs, $\bar{s_u}=\frac{1}{N(N-1)}\sum_{i\neq j}s_{ij}$, and
the average object similarity over all object pairs,
$\bar{s_o}=\frac{1}{M(M-1)}\sum_{\alpha \neq \beta}s_{\alpha\beta}$.
The connections for both users and objects are considered to be
highly clustered since $C_u\gg \bar{s_o}$ and $C_o\gg \bar{s_u}$.
The clustering-degree correlations for users are reported in Fig. 6.
For audioscrobbler.com, a remarkable negative correlation for
small-degree users is observed. Actually, $C_u(k)$ decays in an
exponential form for small $k$. This result agrees with our daily
experience that a heavy listener generally has broader interests of
music\footnote{In the statistical level, the collaborative
clustering coefficient reflects the diversity of a user's tastes:
the higher coefficient corresponds to the narrower tastes.}. In
contrast, for del.icio.us a weakly positive correlation is observed
for small-degree users. One reason for the difference between
audioscrobbler.com and del.icio.us is that the collections in
audioscrobbler.com only reflect the particular tastes of music,
while the collections of URLs contain countless topics wherein music
is just a very small one. In audioscrobbler.com, collections of a
heavy listener (i.e., large-degree user) usually consist of several
music genres, each of which contains a considerable number of music
groups, while most of the music groups collected by a small-degree
user belong to one genre. However, in del.icio.us, even for a
very-small-degree user, his/her few collected URLs can be of highly
diverse topics. Therefore, for del.icio.us, one can not infer that a
small-degree user has limited interests. In addition, collections of
music groups are mainly determined by personalized interests, while
we have checked that in del.icio.us, many bookmarks are less
personalized, that is, they can not well reflect the personal
interests of users. For example, online tools like translators and
search engines, and information services webs like the train
schedules and air ticket centers are frequently collected. However,
till now, we are not fully understood the origins of those
nontrivial correlations, a future exploration making use of
content-based or topic-based analysis on the URLs may provide a
clearer picture.

Figure 7 reports the clustering-degree correlations for objects. For
the lower-degree objects, a negative correlation between the object
collaborative clustering coefficient and the object degree is
observed, which disappears at about the average object degree. This
result suggests that the unpopular objects (i.e., small-degree
objects) may be more important than indicated by their degrees,
since the collections of unpopular objects can be considered as a
good indicator for the common interests--it is not very meaningful
if two users both select a popular object, while if a very unpopular
object is simultaneously selected by two users, there must be some
common tastes shared by these two users. In fact, the empirical
result clear shows that the users commonly collected some unpopular
objects have much higher similarity to each other than the average.
The information contained by those small-degree objects, usually
having little effect in previous algorithms, may be utilized for
better community detection and information recommendation.

\section{Conclusion and Discussion}
Today, the exploding information confronts us with an information
overload: we are facing too many alternatives to be able to find out
what we really need. The collaborative filtering web sites provide a
promising way to help us in automatically finding out the relevant
objects by analyzing our past activities. In principle, all our past
activities can be stored in the user-object networks (maybe in a
weighted manner), which play the central role in those online
services. This Letter reports the empirical analysis of two
user-object networks based on the data downloaded from
audioscrobbler.com and del.icio.us. We found that all the
object-degree distributions are power-law while the user-degree
distributions obey stretched exponential functions, which refines
the previous results \cite{Lambiotte2005b}. For both users and
objects, the connections display disassortative mixing patterns, in
accordance with the observations in user-movie networks
\cite{Grujic2008,Grujic2009}. We proposed a new index, named
collaborative clustering coefficient, to quantify the clustering
behavior based on the collaborative selection. The connections for
both users and objects are considered to be highly clustered since
the collaborative clustering coefficients are much larger than the
corresponding background similarities.

A problem closely related to the analysis of web-based user-object
bipartite networks is how to recommend objects to users in a
personalized manner \cite{Herlocker2004,Adomavicius2005}. The
empirical results reported in this Letter provide some insights in
the design of recommendation algorithms. For example, as shown in
Fig. 4, the average degree of collected objects is negatively
correlated with the user's degree, and the fresh users tend to
select very popular objects, that is, they have not well established
their personalities and their collections are mostly
popularity-based. This phenomenon gives an empirical explanation of
the so-called \emph{cold-start problem} \cite{Schein2002}, namely
the personalized recommendations to the very-small-degree users are
often inaccurate. In addition, if we compare the significance of the
user collaborative clustering coefficient, $C_u/\bar{s_o}$, and the
significance of the object collaborative clustering coefficient,
$C_o/\bar{s_u}$, we will find that for both audioscrobbler.com and
del.icio.usm, the former (268.07 and 72.84) are much larger than the
latter (4.11 and 6.79). Therefore, the fact that some users have
commonly selected an object does not imply that they are much more
similar to each other than two random users, however the objects
selected by a user are statistically much more similar to each other
than two random objects. The collaborative filtering techniques have
two categories in general \cite{Herlocker2004,Adomavicius2005}: one
is user-based, which recommends to the target user the objects
collected by the users sharing similar tastes; the other is
object-based, which recommends the objects similar to the ones the
target user preferred in the past. The comparison between
$C_u/\bar{s_o}$ and $C_o/\bar{s_u}$ indicates that the object-based
collaborative filtering will perform better, and such a kind of
comparison can be considered as a helpful evidence before the choice
between any user-based and object-based algorithms
\cite{Sarwar2001}. Furthermore, the clustering-degree correlations
reported in Fig. 7 suggest that the small-degree objects actually
play a more significant role than indicated by their degrees. In
fact, we have already demonstrated that to emphasize the impacts of
small-degree objects can remarkably enhance the recommendation
algorithms' accuracies \cite{Zhou2008,Liu2009}. We think the further
in-depth analysis of information contained by the small-degree
objects can find its applications in the design of more efficient
and accurate recommendation algorithms.

\acknowledgements We acknowledge the valuable suggestions and
comments from Bosiljka Tadic and Renaud Lambiotte. This work was
partially supported by Swiss National Science Foundation (grant
no.~200020-121848), the National Natural Science Foundation of China
under Grant Nos. 60973069 and 90924011. M.S.S. acknowledges the
China Postdoctoral Science Foundation under Grant No. 20080431273
and the Sino-Swiss Science and Technology Cooperation (SSSTC)
Project EG 20-032009. T.Z. acknowledges the National Natural Science
Foundation of China under Grant Nos. 60744003 and 10635040.

\end{document}